\documentclass[aps,groupedaddress,nofootinbib]{revtex4}

\usepackage[latin1]{inputenc}
\usepackage{graphicx}
\usepackage{amssymb,amsmath}

\begin{document}

\preprint[{\rightline{KCL-PH-TH-2012-10}}

\title{Asymptotic Analysis of the Boltzmann Equation for Dark Matter Relics}

\author{Carl M. Bender\footnote{Permanent address: Department of Physics,
Washington University, St.~Louis, MO 63130, USA.}}
\email[]{cmb@wustl.edu}
\affiliation{Department of Physics, King's College London, Strand, London WC2R
2LS, UK}

\author{Sarben Sarkar}
\email[]{sarben.sarkar@kcl.ac.uk}
\affiliation{Department of Physics, King's College London, Strand, London WC2R
2LS, UK}

\date{\today}

\begin{abstract}
This paper presents an asymptotic analysis of the Boltzmann equations (Riccati
differential equations) that describe the physics of thermal dark-matter-relic
abundances. Two different asymptotic techniques are used, boundary-layer theory,
which makes use of asymptotic matching, and the delta expansion, which is a
powerful technique for solving nonlinear differential equations. Two different
Boltzmann equations are considered. The first is derived from general
relativistic considerations and the second arises in dilatonic string cosmology.
The global asymptotic analysis presented here is used to find the long-time
behavior of the solutions to these equations. In the first case the nature of
the so-called freeze-out region and the post-freeze-out behavior is explored. In
the second case the effect of the dilaton on cold dark-matter abundances is
calculated and it is shown that there is a large-time power-law fall off of the
dark-matter abundance. Corrections to the power-law behavior are also
calculated.
\end{abstract}

\maketitle

\section{Introduction}
\label{s1}

The thermal history of nonbaryonic dark-matter (DM) species is highly relevant
to the shaping of the universe as we find it today. The existence of DM is based
on evidence at many length scales. At the scale of galactic halos, for example,
DM explains the observed flatness of the rotation curves of spiral galaxies
\cite{R1}. According to observations over the past twelve years, 23\% of the
energy of the universe consists of DM. This number has been obtained by best-fit
analyses of astrophysical data to the Standard Cosmological Model, which is a
Friedmann-Robertson-Walker cosmology involving cold DM as the dominant DM
species. The modern data is based on observations of type-Ia supernovae
\cite{R2}, the cosmic microwave background \cite{R3,R4}, baryon oscillations
\cite{R5}, and weak-lensing data \cite{R6}. It should be stressed that estimates
of the DM abundance depend crucially on the theoretical model that is
considered.

In the absence of dilaton effects from string theory, the evolution of the
appropriately normalized number density $Y(x)$ of a DM species $X$ of mass $m_X$
is governed by the Boltzmann equation
\begin{equation}
Y'(x)=-\lambda x^{-n-2}\left[Y^2(x)-Y_{\rm eq}^2(x)\right],
\label{e1}
\end{equation}
which is a Riccati equation in the dimensionless independent variable $x\equiv
m_X/T$, where $T$ is the temperature. The parameter $\lambda$ is a dimensionless
measure of the scattering of DM particles and is regarded as a large number
$\lambda\gg1$. The integer $n=0,\,1,\,2,\,\ldots$ comes from a partial-wave
analysis of the scattering of DM particles; $n=0$ refers to $S$-wave scattering.
For bosonic remnants the function $Y_{\rm eq}(x)$ is the distribution \cite{R7}
\begin{equation}
Y_{\rm eq}(x)=A\int_0^\infty ds\frac{s^2}{e^{\sqrt{s^2+x^2}}-1},
\label{e2}
\end{equation}
where $A=0.145 g/g_*$, $g$ is the degeneracy factor for the DM species, and
$g_*$ counts the total number of massless degrees of freedom \cite{R8}.

As the universe cools and $x$ increases, the nature of the solution $Y(x)$ to
(\ref{e1}) changes rapidly in the vicinity of a value $x=x_f$, the so-called
{\it freeze-out} point, and as $x\to\infty$ the solution $Y(x)$ approaches the
constant $Y_\infty$, called the {\it relic abundance}. Because a closed-form
analytical solution to this Riccati equation is unavailable, an approximate
heuristic approach is customarily used to treat this Riccati equation: One
approximation is made for $x<x_f$ and another is made for $x>x_f$. The solutions
in the two regions are then patched at $x=x_f$. This approach gives an intuitive
and reasonably accurate determination of $Y_\infty$ and it is widely adopted
\cite{R8}.

However, this splitting into two regions is only a mathematical convenience and
there is really no precise value $x_f$. Because the differential equation
(\ref{e1}) is first order, its solution is completely determined by one initial
condition, namely $Y(0)$. The usual method of splitting (\ref{e1}) into two
approximate first-order equations, which are valid in each of two regions, leads
to two conditions, an initial condition and a patching condition. We feel that
this gives rise to an unsatisfactory mathematical discussion that is prevalent
in the literature. The value of $x_f$, for example, becomes explicitly involved
in the determination of $Y_{\infty}$ when there is no reason for this.

Equation (\ref{e1}) is valid in a general-relativistic framework. However, given
the importance of understanding the current thermal-relic abundance of DM in
theories beyond the standard model of particle physics, we also reexamine here
the modifications of (\ref{e1}) due to string cosmology \cite{R9}. String theory
is widely accepted as a leading candidate for physics beyond the standard model,
and it places a constraint on the types of time-dependent backgrounds in
conformally invariant critical theories. As before, we are interested in eras in
which the temperature $T$ satisfies $m_X>T>T_0$, where $T_0$ is the current
temperature of the universe. String cosmology leads to a rolling dilaton source
in the Boltzmann equation \cite{R10} that describes DM species. Including this
source gives an additional linear term in the Boltzmann equation:
\begin{equation}
Y'(x)=-\lambda x^{-n-2}\left[Y^2(x)-Y_{\rm eq}^2(x)\right]+\Phi_0 Y(x)/x,
\label{e3}
\end{equation}
where $\Phi_0$ is a negative dimensionless constant of order 1. 

The purpose of this paper is to study analytically the two Riccati equations
(\ref{e1}) and (\ref{e3}). These equations do not have exact closed-form
solutions. However, because $\lambda$ is a large parameter, one can attempt to
find asymptotic approximations to the solutions. The most direct approach is to
convert these Riccati equations into equations of Schr\"odinger type. When this
transformation is applied to (\ref{e1}), we obtain
\begin{equation}
v''(x)-\frac{n(n+2)}{4x^2}v(x)-\lambda^2x^{-2n-4}Y_{\rm eq}^2(x)v=0.
\label{e4}
\end{equation}
Now, if we set $n=0$, we obtain the standard time-independent Schr\"odinger
equation in which $1/\lambda$ plays the role of $\hbar$. While it is possible to
perform a local analysis of this equation for small $x$ and for large $x$, it
is not easy to use WKB analysis to find a global asymptotic approximation
because the equation is singular at $x=0$ and there is a turning point at $x=
\infty$.

Thus, in this paper we will use two other powerful asymptotic methods from which
we can extract global information. The first method is boundary-layer analysis.
This asymptotic technique, which has been used to solve approximately the
equations of fluid mechanics, gives very accurate results, and it has the
physical advantage of treating freeze-out as a boundary-layer region, very much
like the boundary between two fluids. The second technique, known as the {\it
delta expansion} \cite{R11}, is particularly well-suited to study the transition
from the equilibrium region to the large-$x$ behavior of the solutions without
the necessity of finding approximations to the Boltzmann equation in different
epochs. We will see that the presence of a source in (\ref{e3}) gives a solution
for $Y(x)$ in (\ref{e3}), whose qualitative behavior is significantly different
from the solution for $Y(x)$ in (\ref{e1}).

This paper is organized as follows. In Sec.~\ref{s2} we summarize the derivation
of the Boltzmann equations (\ref{e1}) and (\ref{e3}). In Sec.~\ref{s3} we apply
boundary-layer analysis to study (\ref{e1}) and (\ref{e3}). Next, in
Sec.~\ref{s4} we describe the delta expansion and then use it to study the
approximate behaviors of (\ref{e1}) and (\ref{e3}). Finally, in Sec.~\ref{s5} we
give some brief concluding remarks.

\section{Derivation of the Boltzmann equations}
\label{s2}

In this section we review the derivation of the two Boltzmann equations
(\ref{e1}) and (\ref{e3}).

\subsection{Derivation of (\ref{e1})}
\label{ss2a}

In the hot early universe DM particles interact with themselves and with other
particles. Particle species are assumed to react rapidly enough to maintain
equilibrium. However, the universe expands and cools throughout its history. The
timescale associated with this expansion is determined by the Hubble rate $H$.
There is also a timescale $\Gamma$ associated with the scattering cross-section
(that is, an interaction rate per particle). The dynamics of DM particles
depends on the ratio $\Gamma/H$. When $\Gamma/H\gg1$, conditions for equilibrium
hold and $Y(x)$ follows the canonical distribution obtained from equilibrium
statistical mechanics. However, for $\Gamma/H\ll1$ the DM particles are unable
to maintain equilibrium. There is a crossover to freeze-out behavior in which
$Y(x)$ is asymptotically a constant.

Let us consider a two-body scattering process in which particles of species 1
and 2 scatter reversibly into particles of species 3 and 4. The phase-space
distribution function $f_i\left(\vec{r},\vec{p},t\right)$ for the species $i$
gives the number of particles in an infinitesimal region of phase space around
the position $\vec{r}$ and momentum $\vec{p}$: $f_i\left(\vec{r},\vec{p},t\right
)d^3r\,d^3p$. The main bulk quantity of interest is the number density $n_i
\left(\vec{r},t\right)$, which is given by \cite{R8}
\begin{equation}
n_i\left(\vec{r},t\right)=g_i\int\frac{d^3 p}{(2\pi)^3}\,f_i\left(\vec{r},
\vec{p},t\right),
\label{e5}
\end{equation}
where $g_i$ is the degeneracy factor for the $i$th DM species. The evolution of
such a bulk quantity in the universe is given by the Liouville equation (in the
absence of collisions, for simplicity)
\begin{equation}
\frac{df_i}{dt}=L[f_i]\equiv\left(\frac{\partial}{\partial t}+\frac{d\vec{p}}{dt
}\cdot\nabla_{\vec{p}}+\frac{d\vec{r}}{dt}\cdot\nabla_{\vec{r}}\right)f_i=0.
\label{e6}
\end{equation}

The standard Robertson-Walker metric for an isotropic and expanding flat
universe is given by
\begin{equation}
ds^2=-dt^2+a^2(t)\left(dx^2+dy^2+dz^2\right),
\label{e7}
\end{equation}
where $a(t)$ is the scale factor \cite{R12}. The covariant generalization of
(\ref{e6}) is \cite{R8},
\begin{equation}
L\left[f_i\right]=\left(p^\mu\frac{\partial}{\partial x^\mu}-\Gamma_{\nu
\rho}^\mu p^\nu p^\rho\frac{\partial}{\partial p^\mu}\right)f_i=0,
\label{e8}
\end{equation}
where the Christoffel symbol is given by
$$\Gamma_{\nu\rho}^\mu\equiv g^{\alpha\mu}\left(g_{\alpha\nu,\rho}+
g_{\alpha\rho,\nu}-g_{\nu\rho,\alpha}\right)/2.$$
For the metric in (\ref{e7}), isotropy further implies that $f_i\left(\vec{p},\,
t\right)=f_i\left(\left|\vec{p}\right|,t\right)$. For the isotropic case
(\ref{e8}) takes the form
$$L[f(E,t)]=E\frac{\partial f}{\partial t}-\frac{\dot{a}}{a}\left|\vec{p}
\right|^2\frac{\partial f}{\partial E},$$
where $E=\sqrt{\vec{p}^2+m^2}$.

For a two-body collision process the Liouville equation (\ref{e8}) no longer has
a vanishing right side. This equation can then be used to describe the change in
the number density of a given species. For species 1, for example, one gets
\cite{R8}
\begin{eqnarray}
a^{-3}\frac{d\left(n_1a^3\right)}{dt} &=& \int\frac{d^3p_1}{(2\pi)^3 2E_1}\int
\frac{d^3 p_2}{(2\pi)^3 2E_2}\int\frac{d^3 p_3}{(2\pi)^3 2E_3}\int\frac{d^3p_4}
{(2\pi)^3 2E_4}\nonumber\\
&&\quad\times(2\pi)^4\delta^3\left(p_1+p_2-p_3-p_4\right)\delta\left(E_1+E_2-E_3
-E_4\right)|\mathcal{A}|^2\nonumber\\
&&\quad\times\left[f_3 f_4\left(1\pm f_1\right)\left(1\pm f_2\right)-f_1 f_2
\left(1\pm f_3\right)\left(1\pm f_4\right)\right],
\label{e9}
\end{eqnarray}
where the plus sign is used for a bosonic species and the minus sign is used for
a fermionic species. The symbol $\mathcal{A}$ represents the scattering
amplitude for the process $1+2\leftrightarrow 3+4$ and it is a function of the
$p_i$.

If the scattering process is sufficiently fast, $f_i$ can be parametrized by
canonical Fermi-Dirac or Bose-Einstein distributions. For temperatures $T\ll E-
\mu$ the Bose-Einstein and Fermi-Dirac distributions both take the form
\begin{equation}
f(E)\sim e^{\mu/T}e^{-E/T},
\label{e10}
\end{equation}
which implies that quantum statistics are not important. Hence, the
Pauli-blocking and Bose-enhancement are negligible ($f_i\ll 1$), and the
third line of (\ref{e9}) simplifies:
$$f_3 f_4\left(1\pm f_1\right)\left(1\pm f_2\right)-f_1 f_2\left(1\pm f_3\right)
\left(1\pm f_4\right)\sim e^{-(E_1+E_2)/T}\left[e^{(\mu_3+\mu_4)/T}-
e^{(\mu_1+\mu_2)/T}\right],$$
where the relation $E_1+E_2=E_3+E_4$ has been used. Also, combining (\ref{e5})
and (\ref{e10}), we get
$$n_i=g_i e^{\mu_i/T}\int\frac{d^3 p}{(2\pi)^3}e^{-E_i/T}.$$

The equilibrium number density in the absence of a chemical potential is denoted
by $n_i^{(0)}$. Thus, 
\begin{equation}
a^{-3}\frac{d}{dt}\left(n_1 a^3\right)=n_1^{(0)}n_2^{(0)}\langle\sigma v\rangle
\left\{\frac{n_3 n_4}{n_3^{(0)}n_4^{(0)}}-\frac{n_1 n_2}{n_1^{(0)}n_2^{(0)}}
\right\},
\label{e11}
\end{equation}
where the thermally averaged annihilation cross-section $\langle\sigma v\rangle$
is given by
\begin{eqnarray}
\langle\sigma v\rangle &\equiv& \frac{1}{n_1^{(0)}n_2^{(0)}}\int\frac{d^3 p_1}
{(2\pi)^3 2E_1}\int\frac{d^3 p_2}{(2\pi)^3 2E_2}\int\frac{d^3 p_3}{(2\pi)^3 2E_3
}\int\frac{d^3 p_4}{(2\pi)^3 2E_4}e^{-\left(E_1+E_2\right)/T}\nonumber\\
&&\quad\times(2\pi)^4\delta^3\left(p_1+p_2-p_3-p_4\right)\delta\left(E_1+E_2-E_3
-E_4\right)|\mathcal{A}|^2.
\label{e12}
\end{eqnarray}

We now make the standard assumption \cite{R8} that the predominant interaction
of the cold DM species $X$ of mass $m_X$ is $XX\leftrightarrow ll$, where $l$ is
a light particle in equilibrium. As a consequence, in (\ref{e11}) we can replace
$n_1$ and $n_2$ by $n_X$, where $n_X$ is the number density of the species $X$.
Also, we replace and $n_3$ and $n_4$ by $n_l^{(0)}$. The resulting equation is
\begin{equation}
a^{-3}\frac{d}{dt}\left(n_X a^3\right)=
\langle\sigma v\rangle\left[\left(n_X^{(0)}\right)^2-n_X^2\right].
\label{e13}
\end{equation}

We now define $x\equiv m/T$ and note that $dx/x=-dT/T=da/a$ because $T$ scales
as $1/a$. Thus, $\frac{dx}{dt}=Hx$, where the Hubble rate $H\equiv\frac{d}{dt}
\log(a)$. Since the cosmological era for DM production is radiation dominated,
$a(t)\propto\sqrt{t}$. This translates into $H=H_m/x^2$ with $H_m=1.67g_*^{1/2}
m_X^2/m_{\rm Planck}$. It is known theoretically \cite{R8} that $\sigma v\propto
v^{2n}$ with $n=0$ for $s$-wave annihilation and $n=1$ for $p$-wave
annihilation. Since $\langle v\rangle\propto\sqrt{T}$, we have the
parametrization $\langle\sigma v\rangle=\sigma_0x^{-n}$ for $x\geq3$ \cite{R8}.

Finally, we introduce the dependent variable $Y\equiv n_X/T^3\propto n_Xa^3$.
Similarly, we define $Y_{\rm eq}\equiv n_X^{(0)}/T^3$. We then obtain the
Boltzmann equation in (\ref{e1}), where $\lambda\equiv\sigma_0 m_X^3/H_m\propto
m_{\rm Planck}/m_X$, and this explains why $\lambda$ is a large dimensionless
parameter \cite{R13}.

\subsection{Derivation of (\ref{e3})}
\label{ss2b}

String theory can be formulated in nonflat backgrounds, which is necessary when
considering cosmology. Here, we consider the world-sheet sigma-model approach
for dilaton-based cosmologies \cite{R9}. In superstring theory the bosonic part
of the supermultiplet with lowest energy consists of the following massless
states: the graviton $g_{MN}$, the spinless dilaton $\Phi$, and the
antisymmetric spin-one tensor $B_{MN}$. For expanding universes $\Phi$ provides
consistent time-dependent backgrounds. In such backgrounds the string sigma
model on the world sheet $\Sigma$ is given by \cite{R14}
\begin{equation}
S_\sigma=\int_\Sigma\frac{d^2\sigma}{4\pi\alpha'}\left[\sqrt{\gamma}\,
\gamma^{\alpha\beta}g_{MN}(X)\partial_\alpha X^M\partial_\beta X^N+B_{MN}(X)
\epsilon^{\alpha\beta}\partial_\alpha X^M\partial_\beta X^N+\alpha'\sqrt{\gamma}
\Phi(X)R^{(2)}/2\right],
\label{e14}
\end{equation}
where $X^M$ are target space-time coordinates with $M,N=0,\,1,\,\ldots,\,9$,
$\sigma^\alpha$ are the world-sheet coordinates with $\alpha,\beta=0,\,1$,
$\gamma^{\alpha\beta}$ is the world-sheet metric, $\gamma=\left|\det\left(\gamma
^{\alpha\beta}\right)\right|$, $R^{(2)}$ is the Ricci scalar associated with
$\gamma^{\alpha\beta}$, and $\alpha'$ is the string slope. Expanding around a
conformal flat background with the action $S^*$, we can write $S_\sigma$ as
\begin{equation}
S_\sigma=S^{*}+h^i\int_\Sigma d^2\sigma\,V_i,
\label{e15}
\end{equation}
where $h^i$ denotes the background fields $\left\{g_{MN},B_{MN},\Phi \right\}$
and $V_i$ are associated vertex operators \cite{R14}.

Short-distance singularities of the quantum field theory on the world sheet lead
to renormalized couplings $\left\{h_R^i\right\}$ and to dependence on the
renormalization-group scale $\mu$ \cite{R15}. Usually, this results in
nonvanishing $\beta$ functions: $\beta^i\equiv dh_R^i/d\log\mu$. To restore
conformal invariance, these $\beta$ functions must vanish. This leads to
equations of motion satisfied by the background fields. The usual procedure is
to consider an effective target-space action in the string frame that reproduces
the equations of motion:
\begin{equation}
S=-\frac{1}{2\alpha'^4}\int d^{10}x\,\sqrt{G}\,e^{-\Phi}\left[R+(\nabla\Phi)^2+
2\alpha'^4 U(\Phi)-\tilde{H}^2/12\right],
\label{e16}
\end{equation}
where $\tilde{H}^2=H_{\mu\nu\alpha}H^{\mu\nu\alpha}$, $H_{\mu\nu\alpha}\equiv
\partial_\mu B_{\nu\alpha}+\partial_\nu B_{\alpha\mu}+\partial_\alpha B_{\mu\nu
}$, and the potential $U(\Phi)$ has been introduced. With the help of duality
symmetries it is possible to find analytic solutions for the time dependence of
the dilaton field \cite{R9}.

From (\ref{e16}) it can be shown \cite{R10} that in three spatial dimensions the
energy density $\rho$ of the DM species $X$ satisfies
\begin{equation}
\frac{d\rho}{dt}+3H(\rho+p)-\frac{d\Phi}{dt}(\rho-3p)=0.
\label{e17}
\end{equation}
We then assume that the thermal DM species $X$ behaves like dust (that is, $p=
0$) and that the energy density of the DM is given by the simple formula $\rho=
m_X n_X$. Next, in place of the $0$ on the right side of (\ref{e17}), we include
a collision term, which is just the right side of (\ref{e13}):
\begin{equation}
\frac{d}{dt}n_X+3Hn_X-\frac{d\Phi}{dt}n_X=
\langle\sigma v\rangle\left[\left(n_X^{(0)}\right)^2-n_X^2\right].
\label{e18}
\end{equation}
Assuming that matter sources are perfect fluids and requiring scale-factor
duality symmetry, one can show \cite{R9} that up to an additive constant, $\Phi
(t)=\Phi_0\log a(t)$ where $\Phi_0={\rm O}(1)$ and $\Phi_0<0$.

Finally, we make the assumption that the behavior of the DM species is dominated
by radiation so that $a(t)\propto t^{1/2}$ \cite{R8}. As in Subsec.~\ref{ss2a},
we introduce the variables $Y(x)$ and $x$ and obtain the Boltzmann equation
(\ref{e3}).

\section{Boundary-layer solution to (\ref{e1}) and (\ref{e3})}
\label{s3}

In this section we show how to perform a boundary-layer asymptotic analysis of
(\ref{e1}) and (\ref{e3}). The advantage of this analysis is that it provides a
global picture of the cosmological development from the initial time to the
present as described by the Boltzmann equation, and not just the physics of the
equilibrium epoch or of the post-equilibrium epoch alone. It also establishes a
framework to describe in a clear and natural way the region of rapid transition
between these two epochs. A satisfactory description of this crucial transition
is lacking in earlier treatments in the literature because the earlier analysis
used {\it patching} (joining together two solutions to a differential equation
at an arbitrary and fictitious point, which produces an elbow in the solution)
rather than {\it asymptotic matching} \cite{R16}.

One may wonder why an asymptotic procedure as powerful as boundary-layer theory
should be used to solve a first-order ordinary differential equation as simple
as a Riccati equation. A general Riccati equation
$$y'(x)=a(x)y^2(x)+b(x)y(x)+c(x)$$
can be recast as a linear second-order equation,
\begin{equation}
w''(x)-\left[\frac{a'(x)}{a(x)}+b(x)\right]w'(x)+a(x)c(x)w(x)=0,
\label{e19}
\end{equation}
where $y(x)=-\frac{w'(x)}{a(x)w(x)}$.
Furthermore, (\ref{e19}) can be recast as a Schr\"odinger equation
\begin{equation}
v''(x)+\left\{p''(x)/p(x)-[b(x)+a'(x)/a(x)]^2/2+a(x)c(x)\right\}v(x)=0
\label{e20}
\end{equation}
by introducing $v(x)=w(x)/p(x)$, where $p'(x)/p(x)=[b(x)+a'(x)/a(x)]/2$.

The form (\ref{e20}) is often useful for asymptotic WKB analysis but the problem
of freeze-out poses mathematical difficulties. If we apply these transformations
to (\ref{e1}) for the case $n=0$ and use the leading asymptotic forms for
$Y_{\rm eq}(x)$ in Appendices A and B, we obtain the Schr\"odinger equations
\begin{equation}
v''(x)-\lambda^2 x^{-4}\eta^2v(x)=0,
\label{e21}
\end{equation}
where $\eta\equiv 2A\zeta(3)$ for $x\ll1$, and
\begin{equation}
v''(x)-\lambda^2A^2 x^{-1}e^{-2x}v(x)=0
\label{e22}
\end{equation}
for $x\gg1$. The role of $\hbar$ in these equations is played by $1/\lambda$
because $\lambda$ is treated as a large parameter.

The exact general solution of (\ref{e21}) is
\begin{equation}
v(x)=x\left(v_+e^{\lambda\eta/x}+v_-e^{-\lambda\eta/x}\right),
\label{e23}
\end{equation}
where $v_+$and $v_-$ are constants. The approximate general solution to
(\ref{e22}) can be found by using a standard application of WKB \cite{R16}. [A
detailed analysis of (\ref{e22}) for large $x$ is given in Appendix C.] However,
because (\ref{e22}) has a turning point at $x=\infty$ and because (\ref{e21})
has a singularity at $x=0$, it is very difficult to construct a uniform
asymptotic expansion that is valid for all $x$. We show below that
boundary-layer theory overcomes these difficulties.

\subsection{Boundary-layer analysis of (\ref{e1})}
\label{ss3a}

Whenever the highest-derivative term in a differential equation is multiplied by
a small parameter, one can attempt a boundary-layer analysis \cite{R16}. In such
an analysis one identifies an {\it outer} region (or regions) in which the
solution is slowly varying and an {\it inner} or {\it boundary-layer} region (or
regions) in which the solution is rapidly varying. If these regions have an
overlap, one tries to construct a global asymptotic approximation to the
differential equation by performing an asymptotic match of the outer solutions
to the inner solutions.

In boundary-layer form the derivative term in (\ref{e1}) is multiplied by $1/
\lambda$, which is regarded as small ($1/\lambda\ll1$). Thus, we begin by
looking for an outer solution; that is, a solution whose derivative is not
large. To leading order such a solution in the outer region satisfies a {\it
distinguished limit} (an asymptotic balance between two of the three terms in
the differential equation) in which we neglect the derivative term as $\lambda
\to\infty$:
\begin{equation}
Y(x)\sim Y_{\rm eq}(x)\quad(\lambda\to\infty).
\label{e24}
\end{equation}
Since $Y(x)$ is well approximated by $Y_{\rm eq}(x)$ in this region, we call
this outer region the {\it thermal-equilibrium} region.

To higher order, we seek a series expansion of this thermal-equilibrium outer
solution as a formal power series in inverse powers of $\lambda$:
\begin{equation}
Y^{\rm thermal-equilibrium}(x)\sim\sum_{k=0}^\infty\lambda^{-k}Y_k^{\rm
thermal-equilibrium}(x).
\label{e25}
\end{equation}
Substituting this series into (\ref{e1}) and collecting powers of $1/\lambda$
yields the higher-order terms in the outer series. For example, to first order
we get
\begin{equation}
Y_1^{\rm thermal-equilibrium}(x)=-\frac{1}{2}x^{n+2}\frac{d}{dx}\log\left[Y_{\rm
eq}(x)\right].
\label{e26}
\end{equation}

We must now determine the extent of the thermal-equilibrium region. We know from
Appendix A that for large $x$, $x\gg1$, the asymptotic behavior of $Y_{\rm eq}
(x)$ is given by
\begin{equation}
Y_{\rm eq}(x)\sim Ae^{-x}x^{3/2}\qquad(x\to\infty).
\label{e27}
\end{equation}
Thus, for large $x$ in the outer region
\begin{equation}
Y_0^{\rm thermal-equilibrium}(x)\sim Ae^{-x}x^{3/2}\quad{\rm and}\quad Y_1^{\rm
thermal-equilibrium}(x)\sim\frac{1}{2}x^{n+2}.
\label{e28}
\end{equation}
Hence, the second term in the outer series is no longer small compared with the
first term when
\begin{equation}
x\sim\log(2A\lambda)-(n+1/2)\log(x).
\label{e29}
\end{equation}
We will call the solution to this asymptotic relation the so-called {\it
freeze-out} value $x_{\rm f}$:
\begin{equation}
x_{\rm f}\sim\log(2A\lambda)-(n+1/2)\log\left(x_{\rm f}\right).
\label{e30}
\end{equation}
Note that if we take $\lambda\approx 10^{14}$ and $A\approx0.00145$, we see that
the outer asymptotic approximation ceases to be valid when $x$ exceeds the
approximate numerical value
\begin{equation}
x_{\rm f}\approx25.
\label{e31}
\end{equation}

Equation (\ref{e29}) defines the upper asymptotic limit of the
thermal-equilibrium region. However, it is important to emphasize here that {\it
freeze-out does not occur at a point}; $x_{\rm f}$ should not be viewed as a
number but rather as a large range of values of $x$ all satisfying the
asymptotic relation (\ref{e29}):
\begin{equation}
x\sim x_{\rm f}\quad(\lambda\to\infty).
\label{e32}
\end{equation}

A second possible distinguished limit of (\ref{e1}) could in principle consist
of an asymptotic balance between the left side and the second term on the right
side. However, this distinguished limit is inconsistent and must be rejected
because we are led to a contradiction: If we solve the resulting equation, we
find that for large $\lambda$ the first term on the right side is in fact {\it
not negligible} compared with the second term .

A third distinguished limit of (\ref{e1}) occurs when $x$ is so large that the
contribution of the equilibrium term $Y_{\rm eq}^2(x)$ is negligible. In this
case, the left side is asymptotic to the first term on the right side:
\begin{equation}
Y'(x)\sim -\lambda x^{-n-2}Y^2(x)\qquad(x\gg1).
\label{e33}
\end{equation}
In this second outer region, which we will call the {\it post-freeze-out}
region, the solution $Y^{\rm post-freeze-out}(x)$ to (\ref{e33}) is
\begin{equation}
Y^{\rm post-freeze-out}(x)\sim\frac{1}{1/C-\lambda x^{-n-1}/(n+1)},
\label{e34}
\end{equation}
where $C$ is a constant of integration to be determined. Note that this solution
is consistent and valid when $x\gg1$ because $Y_{\rm eq}(x)$ is exponentially
small when $x\gg1$. Note also that as $x\to\infty$, $Y^{\rm post-freeze-out}(x)$
approaches the limiting value $C$. Thus, $C$ represents the long-time limiting
value of the relic abundance.

The physical process of freeze-out can be recast in mathematical terms as a
process that occurs in an inner region (or boundary layer), which we treat as a
time interval that is comparatively short relative to the time intervals of
the two outer regions, the thermal-equilibrium region and the post-freeze-out
region. We begin the analysis of the freeze-out boundary layer by determining
the size of this region. To do so, we introduce the {\it inner variable} $X$:
\begin{equation}
x=x_{\rm f}+\kappa X.
\label{e35}
\end{equation}
We regard $|X|$ as a variable that may get large compared to $1$, say as large
as $X_{\rm max}$, but $X$ is still small compared with $\lambda$. Thus, since
$\kappa$ is expected to be a small parameter roughly of order $1/\lambda$, the
boundary layer is narrow because it extends roughly from $x_{\rm f}-\kappa
X_{\rm max}$ to $x_{\rm f}+\kappa X_{\rm max}$.

Making the change of variables (\ref{e35}), from which we get
\begin{equation}
\frac{d}{dx}=\frac{1}{\kappa}\frac{d}{dX},
\label{e36}
\end{equation}
and treating $\kappa X$ as small compared with $x_{\rm f}$, we find that
(\ref{e1}) becomes
\begin{equation}
\frac{1}{\kappa}\mathcal{Y}'(X)=-\lambda x_{\rm f}^{-n-2}\left[\mathcal{Y}^2(X)-
A^2x_{\rm f}^3e^{-2x_{\rm f}}\right],
\label{e37}
\end{equation}
where $\mathcal{Y}(X)=Y(x)$. A consistent dominant balance in this equation is
achieved if we take
\begin{equation}
\kappa=x_{\rm f}^{n+2}/\lambda,
\label{e38}
\end{equation}
and if we make this choice, we must neglect the second term on the right side
because it is of order $\lambda^{-2}$ compared with the first term on the right
side. This gives the simple inner differential equation
\begin{equation}
\mathcal{Y}'(X)=-\mathcal{Y}^2(X),
\label{e39}
\end{equation}
whose solution is
\begin{equation}
\mathcal{Y}(X)=\frac{1}{X+D},
\label{e40}
\end{equation}
where $D$ is an integration constant.

To complete the boundary-layer analysis, we must match the two outer solutions
to this boundary-layer solution. In order to perform the asymptotic match, we
re-express the outer solutions in terms of the inner variable $X$ and then
carry out an asymptotic approximation valid for small $\kappa$ to these
asymptotic approximations.

Let us look first at the outer solution in the post-freeze-out region:
\begin{equation}
Y^{\rm post-freeze-out}(X)\sim\frac{1}{1/C-\lambda \left(x_{\rm f}+\kappa X
\right)^{-n-1}/(n+1)},
\label{e41}
\end{equation}
which simplifies to
\begin{equation}
Y^{\rm post-freeze-out}(X)\sim\frac{1}{X+\frac{1}{C}-\frac{\lambda}{(n+1)\left(
x_{\rm f}\right)^{n+1}}}.
\label{e42}
\end{equation}
The coefficient of $X$ in the denominator is $1$, which agrees exactly with the
coefficient of $X$ in the inner solution (\ref{e40}). Thus, we have achieved an
asymptotic match, and the matching condition relates the constants $C$ and $D$:
\begin{equation}
D=\frac{1}{C}-\frac{\lambda}{(n+1)x_{\rm f}^{n+1}}.
\label{e43}
\end{equation}

Next, we match the boundary-layer solution in (\ref{e40}) to the outer solution
(\ref{e25}) in the thermal-equilibrium region. To do so, we must re-express the
outer solution in (\ref{e25}) in terms of the inner variable $X$. Although we
are matching to just one term of the inner freeze-out solution, it is essential
that we take the first {\it two} terms in the outer thermal-equilibrium series,
and not just the first term, because we have shown that as we approach the
freeze-out region, the first two terms in the outer solution become comparable
in size. Thus, we include a factor of two in the asymptotic behavior
\begin{eqnarray}
Y^{\rm thermal-equilibrium}(x)&\sim&2Ax^{3/2}e^{-x}\nonumber\\
&\sim&2A\left(x_{\rm f}+\kappa X\right)^{3/2}e^{-x_{\rm f}}e^{-\kappa X}
\nonumber\\
&\sim& \frac{1}{X+\frac{\lambda}{x_{\rm f}^{n+2}}}.
\label{e44}
\end{eqnarray}
Because the coefficient of $X$ in this behavior is $1$, we obtain once again a
perfect asymptotic match to the inner freeze-out solution in (\ref{e40}).
This allows us to determine the value of the constant $D$:
\begin{equation}
D=\lambda x_{\rm f}^{-n-2}.
\label{e45}
\end{equation}
Finally, combining this result with (\ref{e43}), we obtain the value of $C$:
\begin{equation}
C=\frac{(n+1)x_{\rm f}^{n+2}}{\lambda\left(n+1+x_{\rm f}\right)},
\label{e46}
\end{equation}
which is our result for the thermal-relic abundance. For $x_{\rm f}$ large
compared with $n+1$ this is in close agreement with the value $(n+1)x_{\rm f}^{
n+1}/\lambda$ given in Ref.~\cite{R8}.

\subsection{Boundary-layer analysis of (\ref{e3})}
\label{ss3b}

The arguments given in Subsec.~\ref{ss3a} apply to a modified version of
(\ref{e3}). We modify (\ref{e3}) as follows. If we let $\varphi=\left|\Phi_0
\right|$, then the substitution 
$$Z(x)=Y(x)x^\varphi$$
reduces (\ref{e3}) to the simpler Riccati equation
\begin{equation}
Z'(x)=-\lambda x^{-n-2}\left[x^{-\varphi}Z^2(x)-x^\varphi Y_{\rm eq}^2(x)\right]
.
\label{e47}
\end{equation}
The advantage of this equation over (\ref{e3}) is that there are only three
rather than four terms, and thus it is easier to identify a dominant balance.

We can now analyze (\ref{e47}) using the procedure adopted in the previous
subsection. In the left outer region (the thermal-equilibrium region) we have
\begin{equation}
Z_0^{\rm thermal-equilibrium}(x)\sim Ae^{-x}x^{\varphi+3/2}\quad{\rm and}\quad
Z_1^{\rm thermal-equilibrium}(x)\sim\frac{1}{2}x^{\varphi+n+2}.
\label{e48}
\end{equation}
From this result we deduce that the freeze-out value $x_{\rm f}$ is given by
$$x_{\rm f}\sim\log(2A\lambda)-(n+1/2)\log\left(x_{\rm f}\right),$$
which is identical to the result in (\ref{e30}). This result shows that to
leading order in $1/\lambda$ the freeze-out temperature is independent of
$\Phi_0$; that is, the location of the freeze-out region is only weakly affected
by the presence of a dilaton.

Next we discuss the right outer region (post-freeze region). The analog of
(\ref{e34}) is
\begin{equation}
Z^{\rm post-freeze-out}(x)\sim\frac{1}{1/C-\lambda x^{-n-1-\varphi}/(n+1+
\varphi)},
\label{e49}
\end{equation}
where $C$ is a constant of integration to be determined by asymptotic matching.
As before, $C$ describes the long-term-abundance behavior. However, when $x$ is
large compared with the freeze-out temperature $\left(x\gg x_{\rm f}\right)$,
$Y(x)$ does not approach a constant. Rather,
\begin{equation}
Y(x)\sim x^{-\varphi}Z(x)\sim x^{-\varphi}C\quad(x\to\infty).
\label{e50}
\end{equation}

In the freeze-out boundary-layer region we again make the change of
variable in (\ref{e35}),
$$x=x_{\rm f}+\kappa X,$$
where the inner variable $X$ may become large compared to $1$, but it is still
small compared with $\lambda$. Thus, since $\kappa$ is expected to be a small
parameter of order $1/\lambda$, the boundary layer is narrow as before. A
consistent dominant-balance gives the value 
\begin{equation}
\kappa=x_{\rm f}^{n+2+\varphi}/\lambda.
\label{e51}
\end{equation}
The inner differential equation then has the form
\begin{equation}
\mathcal{Z}'(X)=-\mathcal{Z}^2(X),
\label{e52}
\end{equation}
where $\mathcal{Z}(X)=Z(x)$. The solution to (\ref{e52}) is
\begin{equation}
\mathcal{Z}(X)=\frac{1}{X+D},
\label{e53}
\end{equation}
where $D$ is an integration constant. This is the analog of (\ref{e40}).

An asymptotic match of the right outer solution to the boundary-layer
solution produces the relation between the constants $C$ and $D$,
\begin{equation}
D=\frac{1}{C}-\frac{\lambda}{(n+1+\varphi)x_{\rm f}^{n+1+\varphi}},
\label{e54}
\end{equation}
which is the analog of (\ref{e43}). Finally, by matching the left outer
solution to the boundary-layer solution, we obtain the value of $C$:
\begin{equation}
C=\frac{(n+1+\varphi)x_{\rm f}^{n+2+\varphi}}{\lambda\left(n+1+\varphi+x_{\rm f}
\right)}.
\label{e55}
\end{equation}

In conclusion, we find that, due to the presence of a dilation, the
thermal-relic abundance in (\ref{e50}) remains {\it time dependent}; it vanishes
as $x\to\infty$ and does not approach a constant. Note also that if we eliminate
the effect of the dilaton by allowing $\Phi_0$ to approach $0$, the results in
(\ref{e50}) and (\ref{e55}) smoothly reduce to that in Subsec.~\ref{ss3a}).

\section{Application of the Delta expansion to (\ref{e1}) and (\ref{e3})}
\label{s4}

In this section we show how to apply the delta expansion to (\ref{e1}) and
(\ref{e3}). We begin with a brief summary of the delta-expansion technique.

\subsection{Summary of the delta expansion}
\label{ss4a}

The delta expansion is an unconventional perturbative technique for solving
nonlinear problems. It was first introduced to treat nonlinear aspects of
quantum field theory \cite{R17}. To prepare for applying it to the Boltzmann
equations (\ref{e1}) and (\ref{e3}), in this subsection we give a brief review
of the delta expansion.

The theme of the delta expansion is to introduce a parameter $\delta$ as a
measure of the nonlinearity of a problem; that is, the departure of the problem
from a corresponding linear problem. We then treat $\delta$ as small ($\delta
\ll1$), and solve the problem perturbatively by expanding about the linear
problem obtained by setting $\delta=0$. The basic ideas of the delta expansion
are explained in Ref.~\cite{R11}.

To illustrate the delta expansion, we consider the Thomas-Fermi nonlinear
boundary-value problem
\begin{equation}
y''(x)=[y(x)]^{3/2}/\sqrt{x},\qquad y(0)=1,~y(+\infty)=0.
\label{e56}
\end{equation}
This problem is extremely difficult and no closed-form analytical solution is
known. We introduce the parameter $\delta$ in the exponent of the nonlinear term
of the differential equation and consider the one-parameter family of problems
\begin{equation}
y''(x)=y(x)[y(x)/x]^\delta,\qquad y(0)=1,~y(+\infty)=0,
\label{e57}
\end{equation}
where we treat $\delta$ as a small perturbation parameter. The solution to the
unperturbed ($\delta=0$) linear problem is $y_0(x)=e^{-x}$, and we use $y_0(x)$
as the first term in the delta expansion of the solution to the nonlinear
problem (\ref{e57}):
\begin{equation}
y(x)=\sum_{k=0}^\infty\delta^k y_k(x).
\label{e58}
\end{equation}
Finally, we recover the solution to the original Thomas-Fermi problem by
setting $\delta=1/2$. Typically, only very few terms are needed in the delta
expansion to recover accurate numerical results. Furthermore, the accuracy
of the delta expansion can by accelerated by using Pad\'e techniques to sum
the delta expansion. In the case of the Thomas-Fermi problem a $(2,1)$-Pad\'e
approximant has a numerical error of about 1\%.

As a second example, consider the quintic polynomial equation
$$x^5+x-1=0,$$
which cannot be solved by quadrature. The real root of this equation is
$x=0.75487767\ldots$. Introducing the perturbation parameter $\delta$, we obtain
the equation
$$x^{1+\delta}+x=1.$$
We then seek a perturbation series of the form
\begin{equation}
x(\delta)=c_0+c_1\delta+c_2\delta^2+c_3\delta^3+\ldots
\label{e59}
\end{equation}
whose first term is $c_0=1/2$. The radius of convergence of the delta series
(\ref{e59}) is $1$, and therefore it diverges at $\delta=4$. However, a $(3,
3)$-Pad\'e approximant has a numerical error of $0.05\%$ and a $(6,6)$-Pad\'e
approximant has a numerical error of $0.00015\%$.

\subsection{Delta expansion for (\ref{e1})}
\label{ss4b}

To apply the delta expansion to (\ref{e1}), we insert the parameter $\delta$
in such a way that when $\delta=1$ we recover (\ref{e1}):
\begin{equation}
Y'(x)=-\lambda x^{-n-2}\left(Y-Y_{\rm eq}\right)(Y+Y_{\rm eq})^\delta.
\label{e60}
\end{equation}
There are, of course, many ways to insert the parameter $\delta$, but the
advantage of (\ref{e60}) is that the solution to the unperturbed linear problem
obtained by setting $\delta=0$ is qualitatively similar to the solution to
(\ref{e1}), which we have already investigated in Sec.~\ref{s3}. In particular,
when $\delta=0$, $Y(x)$ behaves like $Y_{\rm eq}(x)$ for small $x$, undergoes a
transition as $x$ increases, and then approaches a constant as $x\to\infty$.

Following the usual delta-expansion procedure, we represent $Y(x)$ as a series
in powers of $\delta$,
$$Y(x)=\sum_{k=0}^\infty y_k(x)\delta^k,$$
and then substitute this series into (\ref{e60}). Comparing powers of $\delta$,
we obtain a sequence of {\it inhomogeneous} differential equations for $y_k$:
\begin{equation}
y_k'(x)+\lambda x^{-n-2}y_k(x)=h_k(x)\quad(k=0,\,1,\,2,\,\ldots),
\label{e61}
\end{equation}
where
\begin{eqnarray}
h_0(x)&=&\lambda x^{-n-2}Y_{\rm eq}(x),\nonumber\\
h_1(x)&=&\lambda x^{-n-2}\left[Y_{\rm eq}(x)-y_0(x)\right]\log\left[Y_{\rm eq}
(x)+y_0(x)\right],\nonumber\\
h_2(x)&=&\lambda x^{-n-2}\left\{
y_1(x)\frac{Y_{\rm eq}(x)-y_0(x)}{Y_{\rm eq}(x)+y_0(x)}
+\frac{Y_{\rm eq}(x)-y_0(x)}{2}\log^2\left[Y_{\rm eq}(x)+y_0(x)\right]
-y_1(x)\log\left[Y_{\rm eq}(x)+y_0(x)\right]\right\},
\label{e62}
\end{eqnarray}
and so on.

The solution to (\ref{e61}), which is obtained by using the integrating factor
$\exp\left[-\lambda x^{-n-1}/(n+1)\right]$, has the quadrature form
\begin{equation}
y_k(x)=e^{\lambda x^{-n-1}/(n+1)}\int_0^x ds\,e^{-\lambda x^{-n-1}/(n+1)}h_k(s).
\label{e63}
\end{equation}
Because (\ref{e61}) is a first-order equation, its solution contains one
arbitrary constant for each $k$ and this constant is determined by the
requirement that $y_k(0)$ be finite. This requirement fixes the lower endpoint
of integration to be $0$ for all $k$. Note that if we evaluate the integral in
(\ref{e63}), we obtain the results $y_0(0)=Y_{\rm eq}(0)=2A\zeta(3)$ (see
Appendix B), $y_1(0)=y_2(0)=\ldots=0$. As $x$ increases, $y_0(x)$ remains close
to $Y_{\rm eq}(x)$ until $x$ is of order $\lambda$.

We can now express the freeze-out value $Y(\infty)$ as a series in powers of
$\delta$ and then evaluate this series at $\delta=1$. Here, we just calculate
the first term in ths series:
\begin{equation}
y_0(x)=\lambda e^{\lambda x^{-n-1}/(n+1)}\int_0^\infty ds\,s^{-n-2}
e^{-\lambda s^{-n-1}/(n+1)}Y_{\rm eq}(s).
\label{e64}
\end{equation}
Let us evaluate this integral assuming that the parameter $\lambda$ is large.
Since the integrand is exponentially small for small $s$, we may assume that
the only contribution to the integral comes from the region $s\gg1$, and in
this region we may replace $Y_{\rm eq}(s)$ by its asymptotic behavior $As^{3/2}
e^{-s}$ (see Appendix A). We thus obtain
\begin{equation}
y_0(\infty)\sim A\lambda\int_0^\infty ds\,s^{-n-1/2}e^{\phi(s)}\quad(\lambda\to
\infty),
\label{e65}
\end{equation}
where
$$\phi(s)=-s-\frac{\lambda}{n+1}s^{-n-1}.$$

To evaluate (\ref{e65}) we use Laplace's method with a moving maximum
\cite{R16}. We note that the maximum of $\phi(s)$, which occurs when $\phi'(s)=0
$, is at $s_0=\lambda^{1/(n+2)}$. Hence, we introduce the rescaled variable $t$:
$$s=t\lambda^{1/(n+2)}.$$
This gives the integral
\begin{equation}
y_0(\infty)\sim A\lambda^{5/(2n+4)}\int_0^\infty dt\,t^{-n-1/2}e^{\lambda^{
1/(n+2)}\theta(t)}\quad(\lambda\to\infty),
\label{e66}
\end{equation}
where
$$\theta(t)=-t-\frac{1}{n+1}t^{-n-1}.$$
The maximum of $\theta(t)$ occurs at $t=1$, and near this point we have
the quadratic approximation
$$\theta(t)\sim-\frac{n+2}{n+1}-\frac{n+2}{2}(t-1)^2.$$
Thus, evaluating the Gaussian integral, we obtain the result
\begin{equation}
y_0(\infty)\sim\frac{A\sqrt{2\pi}}{\sqrt{n+2}}\lambda^{2/(n+2)}
\exp\left[-\frac{n+2}{n+1}\lambda^{1/(n+2)}\right],
\label{e67}
\end{equation}
which reduces to
\begin{equation}
y_0(\infty)\sim A\lambda\sqrt{\pi}e^{-2\sqrt{\lambda}}
\label{e68}
\end{equation}
when $n=0$. Thus, the delta expansion predicts that at $x=\infty$ the freeze-out
value of $Y(x)$ is exponentially small.

We see from this calculation that the delta expansion gives a simple and 
qualitatively accurate picture of the solution to the Boltzmann equation
(\ref{e1}). However, the prediction in (\ref{e67}) of the relic abundance
$Y_\infty$ is clearly too small and, of course, this is because we have only
kept the leading-order term in the delta expansion. We will see in the next
subsection that if we retain higher powers of $\delta$, the qualitative features
of the solution do not change but the quantitative prediction for the long-time
behavior of $Y(x)$ is improved.

\subsection{Delta expansion for (\ref{e3})}
\label{ss4c}

The delta expansion treatment of (\ref{e3}) parallels that for (\ref{e1}). We
insert the parameter $\delta$ into (\ref{e3}) as follows:
\begin{equation}
Y'(x)=-\frac{\lambda}{x^{n+2}}[Y(x)-Y_{\rm eq}(x)][Y(x)+Y_{\rm eq}(x)]^\delta
-\frac{\phi}{x}Y(x),
\label{e69}
\end{equation}
where $\phi=\left|\Phi_0\right|$. The analog of (\ref{e61}) is then
\begin{equation}
y_k'(x)+\left(\frac{\lambda}{x^{n+2}}+\frac{\phi}{x}\right)y_k(x)=h_k(x)\quad
(k=0,\,1,\,2,\,\ldots).
\label{e70}
\end{equation}
The solution to (\ref{e70}), which is obtained by using the integrating factor
$x^\phi\exp\left[-\lambda x^{-n-1}/(n+1)\right]$, has the quadrature form 
\begin{equation}
y_k(x)=x^{-\phi}\exp\left[\lambda x^{-n-1}/(n+1)\right]\int_0^x ds\,s^\phi
\exp\left[-\lambda s^{-n-1}/(n+1)\right]h_k(s).
\label{e71}
\end{equation}

Using the modified Laplace method again, we obtain for $x\to\infty$ and large
$\lambda$ the asymptotic approximation
\begin{equation}
y_0(x)\sim x^{-\phi}B(\lambda),
\label{e72}
\end{equation}
where the constant $B(\lambda)$ is given by
$$B(\lambda)=A\lambda^{(\phi+2)/(n+2)}\sqrt{\frac{2\pi}{n+2}}\exp\left[-
\lambda^{1/(n+2)}(n+2)/(n+1)\right].$$
This shows that the dilatonic correction to the Boltzmann equation gives a
significant qualitative change in the freeze-out behavior of DM. The magnitude
of the DM abundance is era dependent because its leading behavior for large $x$
is an algebraic decay of the form $x^{-\phi}$. The delta expansion is
qualitatively in agreement with boundary layer theory.

The result in (\ref{e72}) is the analog of (\ref{e67}), and again we see that
while the delta expansion in leading-order gives a good qualitative description
of the solution to the Boltzmann equation, the quantitative prediction for the
coefficient $B(\lambda)$ of $x^{-\phi}$ in the large-$x$ behavior is much too
small. Thus, we extend the result in (\ref{e72}) to first order in $\delta$. The
calculation is a straightforward generalization of the zeroth-order calculation
and the result is
\begin{equation}
y_0(x)+\delta y_1(x)\sim x^{-\phi}B(\lambda)\left\{1-\delta\log[B(\lambda)]+
\delta\frac{\phi}{n+1}\left[\gamma+\log\left(\frac{\lambda}{n+1}\right)\right]
\right\},
\label{e73}
\end{equation}
where $\gamma=0.5772\ldots$ is Euler's constant.

For large $\lambda$, we can ignore all but the $\log[B(\lambda)]$ term, and we
obtain a rough asymptotic behavior, which is a simplified version of
(\ref{e73}):
\begin{equation}
y_0(x)+\delta y_1(x)\sim x^{-\phi}B(\lambda)\left\{1-\delta\log[B(\lambda)]
\right\}.
\label{e74}
\end{equation}
Not surprisingly, the second-order contribution contains a logarithm squared:
\begin{equation}
y_0(x)+\delta y_1(x)\sim x^{-\phi}B(\lambda)\left\{1-\delta\log[B(\lambda)]
+\frac{1}{2}\delta^2\log^2[B(\lambda)]\right\}.
\label{e75}
\end{equation}
In general, the dominant contribution to the coefficient of $\delta^k$ in the
delta expansion is $(-1)^k\log^k[B(\lambda)]/k!$. Thus, if we sum the
approximate delta series to all orders in $\delta$ and set $\delta=1$, the
multiplicative coefficient $B(\lambda)$, which is numerically incorrect because
it is much too small, is exactly canceled. This explains the mechanism by
which the delta expansion and the matched asymptotic analysis become compatible.

\section{Brief concluding remarks}
\label{s5}

We have applied two powerful perturbative techniques, boundary-layer theory and
the delta expansion, to find globally accurate solutions to two different
Boltzmann equations that describe dark-matter abundances in the early universe.
The first Boltzmann equation is based on the standard model of particle physics
and general relativity; the second includes additional effects due to dilatonic 
contributions that arise in string theory. The boundary-layer solution consists
of contributions from three distinct eras, a thermal-equilibrium epoch, a
freeze-out region, and a nonequilibrium relic-abundance epoch, and the global
solution is obtained by the use of asymptotic matching. The delta-expansion
solution does not require the use of asymptotic matching and gives a good
qualitative picture of the behavior in these three epochs, but the results
to low orders in $\delta$ are not as accurate for long times.

We have shown that when dilatonic effects are not included, the
dark-matter-relic abundance approaches a constant for long times, but when
dilatonic effects are included, the relic abundance has a power-law decay
determined by the dilaton coupling.

\acknowledgments
We are grateful to N.~E.~Mavromatos for enlightening discussions. CMB thanks the
U.K.~Leverhulme Foundation and the U.S.~Department of Energy and SS thanks the
U.K.~Science and Technology Facilities Council for financial support.

\appendix
\setcounter{equation}{0}
\def\theequation{A\arabic{equation}}

\section{Large-$x$ behavior of $Y_{\rm eq}(x)$}
\label{a1}

In this Appendix we derive the large-$x$ asymptotic behavior of the equilibrium
distribution $Y_{\rm eq}$ in (\ref{e2}), whose integral representation is given
by
\begin{equation}
Y_{\rm eq}(x)=A\int_{s=0}^\infty ds\frac{s^2}{e^{\sqrt{s^2+x^2}}-1}.
\label{y1}
\end{equation}
When $x\gg1$, we can neglect $-1$ in denominator of the integrand to all orders
in powers of $1/x$ and write
\begin{equation}
Y_{\rm eq}(x)\sim A\int_0^\infty ds\,s^2 e^{-\sqrt{s^2+x^2}}\quad(x\to\infty).
\label{y2}
\end{equation}
The scaling $s=xt$ followed by the change of variables $u=\sqrt{t^2+1}$ then
gives the integral representation
\begin{equation}
Y_{\rm eq}(x)\sim Ax^3\int_{u=1}^\infty du\,e^{-xu}u\sqrt{u^2-1}\quad(x\to\infty
).
\label{y3}
\end{equation}

Watson's lemma \cite{R16} applies directly to the integral (\ref{y3}). The
procedure is first to expand $u\sqrt{u^2+1}$ as a series in powers of $u-1$, 
\begin{equation}
u\sqrt{u^2-1}=\sum_{n=0}^\infty a_n(u-1)^{n+1/2},
\label{y4}
\end{equation}
where
\begin{equation}
a_n=\frac{1}{\sqrt{2\pi}}(-1/2)^n\frac{(n+3/2)\Gamma(n-3/2)}{n!},
\label{y5}
\end{equation}
and then to interchange orders of summation and integration. Integrating term
by term gives the asymptotic series
\begin{equation}
Y_{\rm eq}(x)\sim Ae^{-x}x^{3/2}\frac{1}{\sqrt{2\pi}}
\sum_{n=0}^\infty \frac{\Gamma(n+5/2)\Gamma(n-3/2)}{(-2x)^n n!}
\quad(x\to\infty).
\label{y6}
\end{equation}
Thus, the series begins
\begin{equation}
Y_{\rm eq}(x)\sim Ae^{-x}x^{3/2}\sqrt{\frac{\pi}{2}}\left(1+\frac{15}{8x}+\ldots
\right)\quad(x\to\infty).
\label{y7}
\end{equation}

\setcounter{equation}{0}
\def\theequation{B\arabic{equation}}

\section{Small-$x$ behavior of $Y_{\rm eq}(x)$}
\label{a2}

In this appendix we show how to find the small-$x$ asymptotic behavior of the
integral
\begin{equation}
Y_{\rm eq}(x)=A\int_{s=0}^\infty ds\frac{s^2}{e^{\sqrt{s^2+x^2}}-1}.
\label{z1}
\end{equation}
We begin by substituting $t=\sqrt{s^2+x^2}$. This gives
\begin{equation}
Y_{\rm eq}(x)=A\int_{t=x}^\infty dt\frac{t\sqrt{t^2-x^2}}{e^t-1}=A\int_{t=x
}^\infty dt\frac{t^2}{e^t-1}\left(\sqrt{1-x^2/t^2}-1+1\right)=\mathcal{A}+
\mathcal{B}+\mathcal{C}+\mathcal{D},
\label{z2}
\end{equation}
where
\begin{eqnarray}
\mathcal{A}&=&Y_{\rm eq}(0)=A\int_{t=0}^\infty dt\frac{t^2}{e^t-1}=2A\zeta(3),
\nonumber\\
\mathcal{B}&=&-A\int_{t=0}^x dt\frac{t^2}{e^t-1},\nonumber\\
\mathcal{C}&=&A\int_{t=1}^\infty dt\frac{t^2}{e^t-1}\left(\sqrt{1-x^2/t^2}-1
\right),\nonumber\\
\mathcal{D}&=&A\int_{t=x}^1 dt\frac{t^2}{e^t-1}\left(\sqrt{1-x^2/t^2}-1\right).
\label{z3}
\end{eqnarray}
We now evaluate each of the integrals $\mathcal{B}$, $\mathcal{C}$, and
$\mathcal{D}$, in turn. 

To evaluate $\mathcal{B}$ we expand $t/(e^t-1)$ in a Taylor series, which
converges if $|t|<2\pi$, and integrate term by term:
\begin{equation}
\mathcal{B}=-A\int_0^x dt\,t\sum_{n=0}^\infty\frac{B_n}{n!}t^n
=-A\sum_{n=0}^\infty \frac{B_n}{(n+2)n!}x^{n+2},
\label{z4}
\end{equation}
where $B_n$ is the $n$th Bernoulli number ($B_0=1$, $B_1=-1/2$, $B_2=1/6$, $B_4=
-1/30$, $\ldots$, $B_{2n+1}=0$ for $n\geq1$). So, 
\begin{equation}
\mathcal{B}=-\frac{A}{2}x^2+\frac{A}{6}x^3-\frac{A}{48}x^4+\ldots.
\label{z6}
\end{equation}

To evaluate $\mathcal{C}$ and $\mathcal{D}$ we use the expansion
\begin{equation}
\sqrt{1-a}-1=-\frac{1}{2\sqrt{\pi}}\sum_{n=1}^\infty\frac{\Gamma(n-1/2)}{n!}a^n.
\label{z7}
\end{equation}
Thus, $\mathcal{C}$ becomes
\begin{equation}
\mathcal{C}=-\frac{A}{2\sqrt{\pi}}\int_1^\infty dt \frac{t^2}{e^t-1}
\sum_{n=1}^\infty\frac{\Gamma(n-1/2)}{n!}x^{2n}t^{-2n}=-\frac{A}{2\sqrt{\pi}}
\sum_{n=0}^\infty\frac{\Gamma(n+1/2)}{(n+1)!}x^{2n+2}\int_1^\infty dt
\frac{t^{-2n}}{e^t-1}.
\label{z8}
\end{equation}
Hence,
\begin{equation}
\mathcal{C}=c_2x^2+c_4x^4+{\rm O}\left(x^6\right),
\label{z9}
\end{equation}
where
\begin{equation}
c_2=-\frac{1}{2}\int_1^\infty\frac{dt}{e^t-1}=\frac{1}{2}\log(1-1/e)
\qquad{\rm and}\qquad c_4=-\frac{1}{8}\int_1^\infty\frac{dt}{t^2(e^t-1)}.
\label{z10}
\end{equation}

The interesting contribution comes from $D$. We express $D$ as the double sum
\begin{equation}
\mathcal{D}=-\frac{1}{2\sqrt{\pi}}\sum_{m=0}^\infty\frac{B_m}{m!}\sum_{n=
0}^\infty \frac{\Gamma(n+1/2)}{(n+1)!}x^{2n+2}\int_{t=x}^1 dt\,t^{m-1-2n}.
\label{z11}
\end{equation}
Depending on the values of $m$ and $n$ in the sum, we get different kinds of
terms. For example, logarithm terms appear when (and only when) $m=2n$. Thus,
all the logarithm terms appear in the series
\begin{equation}
\mathcal{D}_{\rm log~terms}=\frac{1}{2\sqrt{\pi}}\sum_{n=0}^\infty \frac{B_{2n}
\Gamma(n+1/2)}{(n+1)!(2n)!}x^{2n+2}\log x
=\frac{1}{2}x^2\log x-\frac{1}{48}x^4\log x+\ldots.
\label{z12}
\end{equation}

Terms of order $x^2$ arise from the upper endpoint of integration in (\ref{z11})
when $n=0$ and for all $m\geq1$ (but not $m=0$ because this gives rise to a log
term, and we have already included this contribution) and they arise from the
lower endpoint of integration when $m=0$ for all $n\geq1$ (but not $n=0$). The
upper endpoint gives
\begin{equation}
\mathcal{D}_{\rm upper,~2}=-\frac{1}{2}x^2\int_{t=0}^1 dt\left(\frac{1}{e^t-1}-
\frac{1}{t}\right)=-\frac{1}{2}\log(1-1/e)x^2.
\label{z13}
\end{equation}
The lower endpoint gives
\begin{equation}
\mathcal{D}_{\rm lower,~2}=x^2\int_{t=0}^1\frac{dt}{t^3}\left(\sqrt{1-
\frac{1}{2}t^2}+t^2-1\right)=\frac{1}{4}x^2-\frac{1}{2}\log(2)x^2.
\label{z14}
\end{equation}

Thus, the result for the expansion of $I_{\rm eq}(x)$ in (\ref{z1}) to order
$x^2$ is
\begin{equation}
I(x)\sim 2\zeta(3)+\left[\frac{1}{2}\log(x/2)-\frac{1}{4}\right]x^2+\ldots\,.
\label{z15}
% verified by Sarben
\end{equation}

We have pursued this calculation to higher order in powers of $x$, and we find
that in (\ref{z15}) the coefficients of $x^3$ and $x^5$ are 0, the coefficient
of $x^4$ is
\begin{equation}
\frac{\gamma}{96}+\frac{\zeta'(-1)}{8}-\frac{1}{128}-\frac{\log(2)}{96}
+\frac{\log(x)}{96}=-0.0297+0.0104\log(x),
\label{z16}
\end{equation}
and the coefficient of $x^6$ is
\begin{equation}
\frac{\gamma}{192}+\frac{\zeta'(-1)}{16}+\frac{979}{268800}+\frac{\pi}{2880}
-\frac{\log(x)}{11520}=-0.0048+0.0000868\log(x),
\label{z17}
\end{equation}
where $\gamma=0.57721$ is Euler's constant.


\begin{thebibliography}{6}

\bibitem{R1} S.~M.~Faber and J.~J.~Gallagher,
Ann.~Rev.~Astron.~Astrophys.~{\bf 17}, 135 (1979).

\bibitem{R2} A.~G.~Riess {\it et al.} [Supernova Search Team Collaboration],
Astron.~J.~{\bf 116}, 1009 (1998);
% "Observational Evidence from Supernovae for an Accelerating Universe and a
% Cosmological Constant"
S.~Perlmutter {\it et al.} [Supernova Cosmology Project Collaboration],
Astrophys.~J.~{\bf 517}, 565 (1999);
% "Measurements of Omega and Lambda from 42 High-Redshift Supernovae"
R.~Amanullah {\it et al.}, Astrophys.~J.~{\bf 716}, 712 (2010).
% "Spectra and Light Curves of Six Type Ia Supernovae at 0.511<z<1.12 and
% the Union2 Compilation"

\bibitem{R3} D.~N.~Spergel \textit{et al.} [WMAP Collaboration],
Astrophys.\ J.\ Suppl.\ \textbf{170}, 377 (2007).
% "Wilkinson Microwave Anisotropy Probe (WMAP) three year results:
% Implications for cosmology"

\bibitem{R4} E.~Komatsu {\it et al.}, arXiv:1001.4538 [astro-ph.CO] and
references therein.
% "Seven-Year Wilkinson Microwave Anisotropy Probe (WMAP) Observations: 
% Cosmological Interpretation" 

\bibitem{R5} D.~J.~Eisenstein {\it et al.} [SDSS Collaboration],
Astrophys.~J.~{\bf 633}, 560 (2005);
% "Detection of the Baryon Acoustic Peak in the Large-Scale Correlation
% Function of SDSS Luminous Red Galaxies"
M.~Tegmark {\it et al.} [SDSS Collaboration], Phys.~Rev.~D {\bf 69}, 103501
(2004).
% "Cosmological parameters from SDSS and WMAP"

\bibitem{R6} L.~Fu {\it et al.}, Astronomy \& Astrophysics {\bf 479}, 9 (2008);
L.~Guzzo {\it et al.}, Nature {\bf 451}, 541 (2008).

\bibitem{R7} K. Huang, {\it Introduction to Statistical Physics} (Taylor and
Francis, London, 2001).

\bibitem{R8} E.~W.~Kolb and M.~S.~Turner, {\it The Early Universe} (Westview,
Boulder, 1994); S.~Dodelson, {\it Modern Cosmology} (Academic, New York, 2003).

\bibitem{R9} M.~Gasperini, {\it Elements of String Cosmology},
(Cambridge University Press, Cambridge, 2007).

\bibitem{R10} A.~B.~Lahanas, N.~E.~Mavromatos, and D.~V.~Nanopoulos, PMC Phys.~A
{\bf 1}, 2 (2007).
% "Dilaton and off-shell (non-critical string) effects in Boltzmann equation
% for species abundances"

\bibitem{R11} C.~M.~Bender, K.~A.~Milton, S.~S.~Pinsky, and L.~M.~Simmons, Jr.,
J.~Math.~Phys.~{\bf 30}, 1447 (1989).
% "A New Perturbative Approach to Nonlinear Problems"

\bibitem{R12} J.~B.~Hartle, {\it Gravity: An Introduction to Einstein's
General Relativity} (Addison-Wesley, New York, 2003).

\bibitem{R13} J.~L.~Feng, Ann.~Rev.~Astron.~Astrophys.~{\bf 48}, 495 (2010).

\bibitem{R14} N.~E.~Mavromatos, {\it Springer Lecture Notes in Physics 592},
eds.~S.~Cotsakis and E.~Papantonopoulos (Springer, New York, 2002), pp.~392-457.
% "String Cosmology"

\bibitem{R15} M.~E.~Peskin and D.~V.~Schroeder, {\it An Introduction to Quantum
Field Theory} (Westview, Boulder, 1995).

\bibitem{R16} C.~M.~Bender and S.~A.~Orszag, {\it Advanced Mathematical Methods
for Scientists and Engineers} (McGraw Hill, New York, 1978).

\bibitem{R17} C.~M.~Bender, K.~A.~Milton, M.~Moshe, S.~S.~Pinsky, and
L.~M.~Simmons, Jr. Phys.~Rev.~Lett.~{\bf 58}, 2615 (1987) and
% "Logarithmic Approximations to Polynomial Lagrangians"
Phys.~Lett.~B {\bf 205}, 493 (1988) and
% "A New Perturbative Approximation Applied to Supersymmetric Quantum Field
% Theory"
Phys.~Rev.~D {\bf 37}, 1472 (1988);
% "Novel Perturbative Scheme in Quantum Field Theory"
C.~M.~Bender and H.~F.~Jones, Phys.~Rev.~D {\bf 38}, 2526 (1988);
% "New Nonperturbative Calculation: Renormalization and the Triviality of
% $\lambda\phi_4^4$ Field Theory"
C.~M.~Bender and K.~A.~Milton, Phys.~Rev.~D {\bf 38}, 1310 (1988).
% "New Perturbative Calculation of the Fermion-Boson Mass Ratio in a
% Supersymmetric Quantum Field Theory"

\end{thebibliography}
\end{document}